\documentclass[cameraready]{Interspeech}
\usepackage{cite}
\usepackage{amsmath,amssymb,amsfonts}
\usepackage{algorithmic}
\usepackage{graphicx}
\usepackage{textcomp}
\usepackage{amsmath,bm}
\usepackage{multirow}
\usepackage{multicol}
\usepackage{xcolor}
\usepackage{verbatim}
\usepackage{arydshln}
\usepackage{adjustbox}

\title{EII-SCL: Harnessing Emotional Inertia for Multimodal Emotion Recognition in Conversation}

\author[orcid=0009-0003-0760-8350]{Zilong}{Huang}
\author[orcid=0000-0001-9133-3000]{Kong Aik}{Lee}
\author{Chong-Xin}{Gan}
\author{Zezhong}{Jin}
\author{Ruichen}{Zuo}
\author[orcid=0000-0001-8854-3760]{Man-Wai}{Mak}

\address{
  Dept. of Electrical and Electronic Engineering,\\The Hong Kong Polytechnic University, Hong Kong SAR, China
}

\email{\{zi-long.huang, kong-aik.lee,  man.wai.mak\}@connect.polyu.hk}

\keywords{Emotional Inertia, Emotion Recognition in Conversation, Multimodal Network, Contrastive Learning}

\usepackage{comment}

\begin{document}

\maketitle

\begin{abstract}
Multimodal emotion recognition in conversation (MERC) achieves accurate predictions by integrating multimodal and contextual information in dialogues.
While current MERC approaches focus on modeling complex contextual dependencies in conversation, they often overlook the impact of contextual emotional inertia in emotion shift, leading to sub-optimal performance. To address this issue, we propose a novel \textbf{E}motional \textbf{I}nertia-\textbf{I}nformed \textbf{S}upervised \textbf{C}ontrastive \textbf{L}earning module (EII-SCL) that informs the contrastive objective by constructing inertia-affected samples within temporal windows, effectively leveraging emotional inertia as a prior while enabling seamless integration with existing MERC models without requiring additional data. Extensive experiments on IEMOCAP and MELD show that our approach consistently outperforms state-of-the-art methods.
\end{abstract}

\section{Introduction}
\label{sec:intro}

\textit{Emotion recognition in conversation} (ERC) is a key technique in applications such as human-computer interaction and intelligent medical care \cite{yadollahi2017current}. Different from traditional emotion recognition, ERC analyzes dialogue turns and assigns one emotion label per utterance in the dialogue \cite{poria2019emotion,mm-nodeformer}. 

The core of ERC lies in modeling contextual dependencies between utterances. Mainstream methods construct dialogue structures through graph neural networks to explicitly capture such a relationship. 
MMGCN~\cite{hu2021mmgcn} extends graph neural networks to multimodal settings, exploring a more effective way of utilizing both multimodal and long-distance contextual information.
AdaIGN~\cite{AdaIGN} uses an adaptive graph selection strategy to balance intra- and inter-modal interaction, better handling the challenge of balancing cross-speaker contextual dependencies. Recent studies have explored advanced representation learning strategies to improve feature discrimination in conversational and speech-related tasks \cite{extradis,extraunet,curriculum,extradenoising,extraren2025bav}. However, most of the contextual modeling works \cite{COLD,FEMI,extraidir,SDT} overlook emotional inertia, making their models unable to leverage such an important psychological behavior of humans to improve ERC performance.

Emotional inertia \cite{inertia1,inertia2} is a psychological concept that describes the resistance of emotions to change. In a conversational environment, an individual's previous emotional state will have a lasting impact on current and future emotions, i.e., the emotional state of a speaker tends to maintain a gradual and smooth transition rather than irregular and drastic changes. Current research generally studies the phenomenon of emotion shift between discourses, such as introducing emotion shift detection modules \cite{emoshift}, utilizing conversation-level curriculum learning based on emotion shift frequency to progressively enhance the model's ability to recognize emotions \cite{yang2022hybrid}, and modeling multiple emotion labels \cite{one-hot, kang2025beyond}. Although these studies have successfully modeled the emotion shift phenomenon, they failed to fully consider the emotional dependence of the same speaker in continuous emotion expressions and emotion transitions, which limits the in-depth understanding of the dynamics of emotional inertia.

To address the above challenges, we propose a module called \textit{emotional inertia-informed supervised contrastive learning} (EII-SCL). 
Specifically, we introduce a supervised contrastive loss, which imposes constraints on multimodal features based on the emotional inertia of the same speaker within a local time window. As a general module, EII-SCL can be integrated into typical MERC models and improve their performance without adding additional data annotations.
\begin{figure*}[t]
\centering
\includegraphics[width=1.0\textwidth]{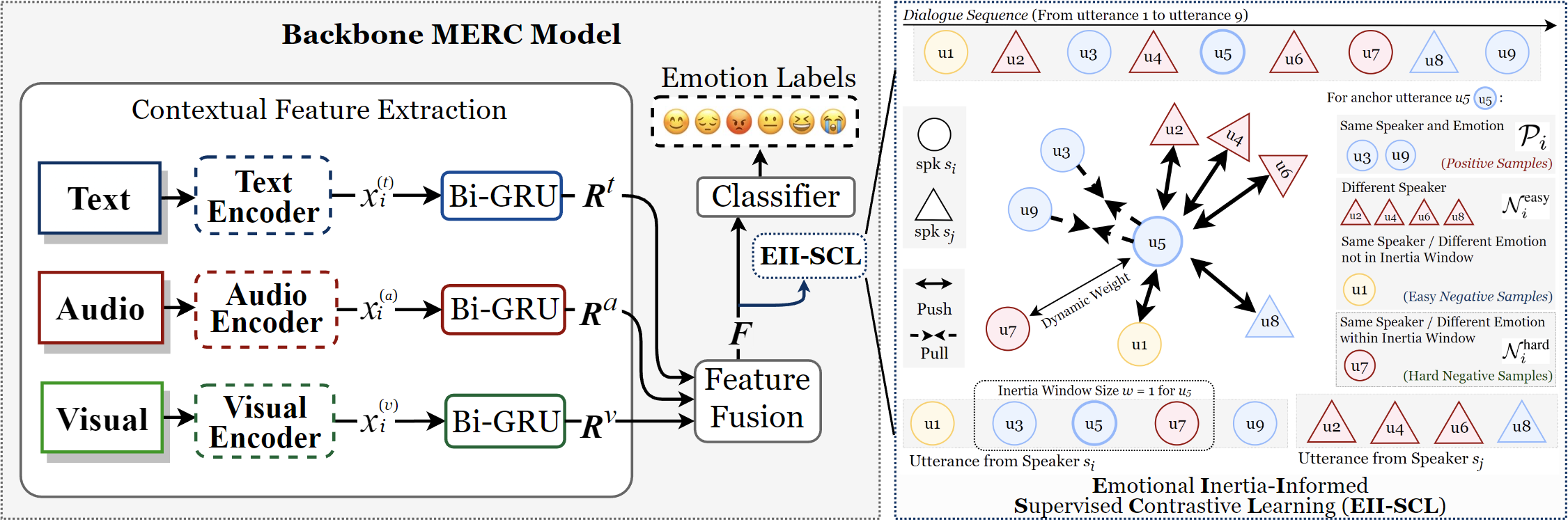}
\caption{The overall architecture of the proposed EII-SCL approach. The backbone MERC model completes emotion prediction through contextual features extraction, multimodal feature fusion, and emotion classification. EII-SCL works with the backbone network to learn more subtle and discriminative feature representations through a carefully designed inertia-informed sampling strategy. The thickness of the lines indicates the intensity of the repulsive force, and different colors of the utterances represent different emotion labels. Solid-line and dashed-line rectangles represent trainable and frozen models, respectively. }
\label{overall}
\end{figure*}
Compared with existing work, our main contributions are reflected in three aspects:

\begin{itemize}
\item This paper proposes an innovative supervised contrastive learning module called EII-SCL. The module constructs hard-negative samples influenced by emotional inertia by jointly leveraging speaker identity, emotional labels, and local time windows, and it designs a dynamic weighting mechanism for these samples to capture the gradual transitions between different emotional states.
\item By rationally selecting positive, easy-negative and hard-negative samples, we analyze and validate the emotional inertia phenomenon on emotion recognition datasets.
\item Our experiments demonstrate that EII-SCL can harness emotional inertia for MERC tasks without adding additional data. A basic MERC model equipped with the EII-SCL module can outperform the current SOTA methods.
\end{itemize}

\section{Methodology}
\subsection{Task Definition}
Consider a dialogue consisting of $k$ temporally ordered utterances, denoted as ${\cal U}=\{u_i\}_{i=1}^k$, where individual utterances have corresponding emotion labels $\{y_i\}_{i=1}^k$ and speaker labels $\{s_i\}_{i=1}^k$. The emotion label ${y_i}$ belongs to a predefined set of emotions ${\cal Y}$. Each utterance ${u_i}$ is accompanied by multimodal data, including a video clip, an audio segment, and a text transcript. Formally, we represent a dialogue as follows:
\begin{equation}
    {\cal U}\times{\cal Y}\times{\cal S}  = \left\{ \{ u_i^{(\delta)} ,{y_i},{s_i}\} \ | \ \delta  \in \{ a,t,v\} ,{y_i}\in{\cal Y}, {s_i} \in {\cal S} \right\}, \label{eq1}
\end{equation}
where $u_i^{(\delta)}$ denotes the modality-specific representation (text, audio, or video) of the $i$-th utterance. The ERC task involves predicting the emotion label ${\hat y_i}$ for each utterance ${u_i}$ in the dialogue, leveraging information from all $k$ utterances.

\subsection{Overall Framework}
As shown in Figure \ref{overall}, the proposed \textit{emotional inertia-informed supervised contrastive learning} (EII-SCL) framework consists of the following components. 
First, a basic MERC model processes the conversation data and generates the final emotion predictions. The EII-SCL module processes the fused multimodal features. Through a supervised contrast loss function, the module improves the perception of emotional inertia of the MERC and feature fusion models through supervised negative samples construction.

\subsection{Backbone MERC Model}
A basic MERC model typically consists of three parts: contextual feature extraction, multimodal feature fusion, and emotion classification.
For each modality, we employ pre-trained encoders to obtain utterance-level representations:
\begin{equation}
    \bm{x}_i^{(\delta)} = f_{\text{enc}}^{(\delta)}(u_i^{(\delta)}),\text{ } \delta \in \{t,a,v\},\text{ }i = 1,\ldots,k,
\end{equation}
where $\delta \in \{t,a,v\}$ and the modality-specific embedding is obtained by averaging the final layer's outputs of the corresponding encoder.
To capture the temporal dependencies within a dialogue, the modality-specific embeddings are processed using bidirectional GRUs:
\begin{equation}
    \bm{R}^{(\delta)} = [\bm{r}_1^{(\delta)},\ldots,\bm{r}_k^{(\delta)}] 
    = \operatorname{Bi\text{-}GRU}(\bm{x}_1^{(\delta)},\ldots,\bm{x}_k^{(\delta)}).
\end{equation}

The contextual representations from different modalities are fused via a feature fusion module (e.g., Transformer-based \cite{chudasama2022m2fnet} or Graph-based \cite{ghosal2019dialoguegcn}):
\begin{equation}
    \bm{F} = [\bm{f}_1,\ldots,\bm{f}_k] 
    = \operatorname{FeatureFusion}[\bm{R}^{(t)},\bm{R}^{(a)},\bm{R}^{(v)}],
\end{equation}
where $\bm{f}_i \in \mathbb{R}^{D_f}$ denotes the fused multimodal embedding of utterance $u_i$.
Each fused embedding $\bm{f}_i$ is fed into a classifier to obtain emotion probabilities $\bm{p}_i$, optimized using the standard cross-entropy loss $\mathcal{L}_{CE}$.
\subsection{Emotional Inertia-Informed Contrastive Learning}
The key to modeling emotional inertia lies in effectively distinguishing between utterances influenced by emotional inertia and those that are not. In this work, we apply the concept of emotional inertia in the context of contrastive learning. Previous supervised contrastive learning methods do not take into account the emotion labels of adjacent utterances, i.e., they only treat utterances with the same emotion label as positive pairs and those with different emotion labels as negative pairs, regardless of their locality in the dialog.

Emotional inertia reflects the temporal persistence and resistance to change of emotional states. This characteristic suggests an important observation: in a dialogue, even when an emotion shift occurs, adjacent utterances from the same speaker often remain semantically similar due to the gradual transition between emotional states, although they are assigned different discrete emotion labels. If a contrastive learning framework ignores such emotion transitions when forming negative pairs, it fails to account for the dynamic influence of emotional inertia, thereby limiting the model’s ability to capture the gradual evolution of emotional states.

To address the aforementioned issue, we redefine the positive and negative sets by incorporating an adaptive inertia window that dynamically adjusts the context scope based on the semantic similarity between utterances. 
The positive pairs $\mathcal{P}_i$ for the $i$-th utterance consist of samples from the same speaker with the same emotion label:
\begin{equation}
\mathcal{P}_i = \{ \bm{f}_j \mid s_j = s_i, y_j = y_i, \forall j \neq i \},
\end{equation}
where $s_i$ and $y_i$ denote the speaker and emotion labels of utterance $u_i$, respectively. 
To adaptively determine the contextual scope of emotional inertia, we estimate the inertia window using an attention mechanism. Specifically, the attention weight between the anchor utterance $i$ and another utterance $j$ from the same speaker is computed as:
\begin{equation}
a_{i,j} =
\frac{\exp\big((\bm{W}_q\bm{f}_i)^\top (\bm{W}_k\bm{f}_j)\big)}
{\sum_{\substack{t \neq i \\ s_t = s_i}}
\exp\big((\bm{W}_q\bm{f}_i)^\top (\bm{W}_k\bm{f}_t)\big)},\text{ } s_j=s_i
\end{equation}
where $\bm{W}_q$ and $\bm{W}_k$ are learnable query and key projection matrices, and $t$ indexes all utterances from the same speaker except the anchor $i$.
The inertia window size is then defined as the attention-weighted temporal distance:
\begin{equation}
\omega_i =
\left\lfloor
\sum_{\substack{j \neq i,\text{ } s_j = s_i}}
a_{i,j}\,|i-j|
\right\rfloor ,
\end{equation}
where $|i-j|$ denotes the index-based temporal distance between the $j^\text{th}$ utterance and $i^\text{th}$ utterance spoken by the same speaker, and $\lfloor \cdot \rfloor$ converts the result to an integer window size. For the negative set, we define the emotional inertia window $\mathcal{W}_i$ as:
\begin{equation}
\mathcal{W}_i = \{ j \mid s_j = s_i, \forall |i-j| \leq {\omega_i} \},
\end{equation}
By using the emotional inertia window, we construct two distinct negative sample sets: $\mathcal{N}_i^{\text{easy}}$ and $\mathcal{N}_i^{\text{hard}}$:
\begin{equation}
\begin{split}
    &\mathcal{N}_i^{\text{easy}} = \{ \bm{f}_j \mid s_j \neq s_i \} \cup \{ \bm{f}_j \mid s_j = s_i, y_j \neq y_i, j \notin \mathcal{W}_i \} \\
    &\mathcal{N}_i^{\text{hard}} = \{ \bm{f}_j \mid s_j = s_i, y_j \neq y_i, j \in \mathcal{W}_i \}.
\end{split}
\end{equation}
The easy-negative set $\mathcal{N}_i^{\text{easy}}$ combines utterances from different speakers with those from the same speaker but different emotion labels and falls outside the inertia window $\mathcal{W}_i$.
The hard-negative set $\mathcal{N}_i^{\text{hard}}$ captures the emotional states of same-speaker utterances with different emotion labels within the inertia window $\mathcal{W}_i$, reflecting the emotional persistence characteristic of emotional inertia. This formulation models emotional inertia by considering speaker-specific temporal windows.

Given the fused embeddings $\{\bm{f}_i\}_{i=1}^k$, we define the emotional inertia-informed contrastive loss as:

\begin{equation}
\begin{split}
    S_{i}^+ &= \sum_{\bm{f}_j \in \mathcal{P}_i}{\exp(\text{cos}(\bm{f}_i,\bm{f}_j)/\tau)}, \\
    S_{i}^- &= \sum_{\bm{f}_t \in \mathcal{N}_i^{\text{easy}}} \exp(\text{cos}(\bm{f}_i,\bm{f}_t)/\tau) \\ 
    &+ \sum_{\bm{f}_h \in \mathcal{N}_i^{\text{hard}}} \underbrace{\left(\frac{1-\text{cos}(\bm{f}_i,\bm{f}_h)}{2}\right)}_{\text{dynamic weight}}\cdot\exp(\text{cos}(\bm{f}_i,\bm{f}_h)/\tau),
\end{split}
\end{equation}
where \text{cos}(,) denotes the cosine similarity and $\tau$ is a temperature parameter. The dynamic weight adjusts the repulsive force based on their similarity to the anchor, thereby avoiding over-penalizing potentially similar utterances while maintaining discriminability. The final contrastive loss for emotional inertia modeling is computed as:
\begin{equation}
\mathcal{L}_{eii} = -\frac{1}{k} \sum_{i=1}^k \log\frac{S_{i}^+}{S_{i}^+ + S_{i}^-}\ .
\end{equation}

\subsection{Joint Training}
The proposed model is optimized with a loss function $\mathcal{L}$ that combines the basic cross-entropy loss $\mathcal{L}_{CE}$ with $\mathcal{L}_{eii}$ for emotional inertia-informed learning. Specifically, EII-SCL collaborates with the Basic MERC model by jointly minimizing the following losses:
\begin{equation}
\mathcal{L} = \mathcal{L}_{CE} + \alpha \mathcal{L}_{eii},
\end{equation}
where $\alpha$ is the coefficient used to balance the two losses. 

\begin{table*}[htbp]
    \setlength{\tabcolsep} {1.7mm}
    \centering
    \caption{Accuracy and weighted-average F1 score (w-F1) compared with other baselines. Best and second-best results are displayed in \textbf{bold} and \underline{underline}, respectively. “MM-TransFormer” means using a simple Transformer encoder as the feature fusion module to fuse the concatenated multimodal features. * indicates our reproduced results from DialogGCN under the multimodal setting (T+A+V).}
    \ninept
    \begin{tabular}{c|c|c|c|c|c|c|c|c|c|c|c}
    \hline
    \multirow{2}{*}{\textbf{Baseline Model}} & \textbf{Proposed} & \multicolumn{8}{c|}{\textbf{IEMOCAP}} & \multicolumn{2}{c}{\textbf{MELD}} \\ \cline{3-12} 
    & \textbf{Year} & \textit{Happy} & \textit{Sad} & \textit{Neutral} & \textit{Angry} & \textit{Excited} & \textit{Frustrated} & \textbf{Acc} & {\textbf{w-F1}} & \textbf{Acc} & {\textbf{w-F1}} \\ \hline
    
    DialogGCN \cite{ghosal2019dialoguegcn} & 2019 & 42.75 & \underline{84.54} & 63.54 & 64.19 & 63.08 &  66.99 & 65.25 & 64.18 & - & 58.10 \\
    MMGCN \cite{hu2021mmgcn} & 2021 & 45.14 & 77.16 & 64.36 & 68.82 & 74.71 & 61.40 & 66.36 & 66.26 & 60.42 & 58.31 \\
    CFN-ESA \cite{li2023cfn} & 2023 & 53.67 & 80.60 & \underline{71.65} & 70.32 & 74.82 & 68.06 & 71.04 & 70.78 & \underline{67.85} & 66.70 \\
    AdaIGN \cite{AdaIGN} & 2024 & 53.04 & 81.47 & 71.26 & 65.87 & 76.34 & 67.79 & - & 70.74 & - & 66.79 \\
    DER-GCN \cite{DER-GCN}  & 2025 & 58.80 & 79.80 & 61.50 & \underline{72.10} & 73.30 & 67.80 & 69.70 & 69.40 & 66.80 & 66.10 \\
    FEMI \cite{FEMI} & 2025 & 60.60 & \textbf{85.55} & 70.92 & 70.98 & 75.13 & 69.00 & 71.97 & \underline{73.53} & 64.88 & 66.41 \\
    \hline
    \hline
    MM-DialogGCN* &   & 73.53 & 74.14 & 67.35 & 71.12 & 73.48 & 71.53 & 71.45 & 71.14 & 67.32 & 66.28 \\
    -with \textbf{EII-SCL} (Proposed) & 2026 & \textbf{76.92} & 65.83 & 69.42 & \textbf{72.24} & \underline{78.42} & \textbf{73.26} & \underline{73.13} & 73.15 & 67.83 & \underline{66.97} \\
    \hline
    MM-TransFormer &   & 73.84 & 67.48 & 70.90 & 71.18 & 75.56 & 72.43 & 72.53 & 72.57 & 67.64 & 66.58 \\
    -with \textbf{EII-SCL} (Proposed) & 2026 & \underline{75.81} & 77.59 & \textbf{72.16} & 69.10 & \textbf{79.09} & \underline{72.79} & \textbf{73.95} & \textbf{74.01} & \textbf{68.19} & \textbf{67.33} \\
    \hline
    \end{tabular}
    \label{tab:results}
    \vspace{-1.9em}
\end{table*}
\section{Experiments}

\subsection{Datasets and Evaluation Metric}

\textbf{IEMOCAP} \cite{busso2008iemocap} is a widely used dataset for emotion recognition in conversation. For partitioning the data, we utilized the commonly used LOSO (Leave-One-Session-Out) strategy. 
\textbf{MELD} \cite{poria2018meld} is a multi-modal, multi-speaker conversational dataset derived from the TV series Friends. To ensure a fair comparison, we followed the predefined train/val/test splits provided by the dataset, ensuring data allocation aligned with \cite{hu2022mm}. 
Similar to other recent studies, we adopt the accuracy and weighted F1-score (w-F1) for evaluation. 

\subsection{Backbone MERC Models and Implementation Details}
    In order to measure the effectiveness of our proposed EII-SCL module, we constructed our backbone model through two representative MERC methods: Transformer-based and Graph-based methods.
    
    The Transformer-based method uses the concatenated features through a transformer encoder to further capture multimodal complementary information, which we refer to as MM-TransFormer:
\begin{equation}
    \bm{F} = [\bm{f}_1,\ldots,\bm{f}_k] = \operatorname{TransEncoder}[\bm{R}^t,\bm{R}^a,\bm{R}^v], \label{TrsEncoder}
\end{equation}
    The Graph-based method refers to the graph construction approach of DialogueGCN \cite{ghosal2019dialoguegcn}. Given that this model primarily focuses on the single-text modality, we extend it to a multimodal framework for a fair comparison. Specifically, we concatenate the multimodal features as the input, construct the graph, and perform feature transformation in a manner consistent with the original configurations, which we refer to as MM-DialogueGCN:
\begin{equation}
    \bm{F} = [\bm{f}_1,\ldots,\bm{f}_k] = \operatorname{DialogueGCN}[\bm{R}^t,\bm{R}^a,\bm{R}^v], \label{DGCN}
\end{equation}
    We integrate the proposed EII-SCL into these two backbone MERC models to evaluate the impact of EII-SCL.
    
    To ensure that methods remain consistent at the feature extraction level, we used the same pre-trained encoder setting to extract features of the three modalities in the backbone MERC models. Specifically, we use RoBERTa \cite{liu2019roberta}, Wav2vec2.0 \cite{baevski2020wav2vec}, and CLIP \cite{radford2021learning} for the text, audio, and visual modalities, respectively. 
    We conducted all experiments on a single NVIDIA RTX 4090 GPU, training for 40 epochs on IEMOCAP and 50 epochs on MELD with a batch size of 13. The model was optimized using Adam with a learning rate of 0.0002 and a dropout rate of 0.3. The temperature $\tau=0.07$ and scaling factors $\alpha=0.02$. Finally, we averaged the weights from the last 10 checkpoints to obtain the final model for evaluation.
\vspace{-1em}
\subsection{Overall Results} 
Table~\ref{tab:results} shows the detailed results on both datasets. After integrating the EII-SCL module, both backbone networks significantly outperform all baseline models. Notably, while the FEMI \cite{FEMI} shows outstanding performance in recognizing “\textit{frustrated}” and “\textit{sad}” emotions, its w-F1 score for “\textit{happy}” is significantly lower (60.60\%), revealing potential emotional label bias in conventional approaches. 

\subsection{Emotional Inertia Analysis in IEMOCAP Dataset}
\begin{figure}[h]    
\centering
\includegraphics[width=0.483\textwidth]{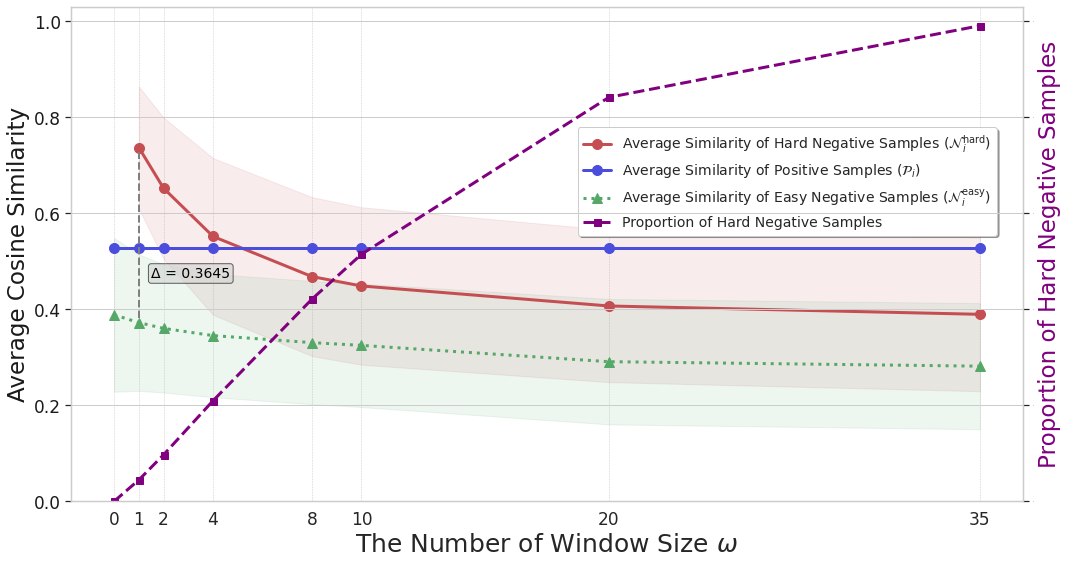} 
\caption{Impact of window size $\omega$ on the average similarity and proportion of negative samples on the IEMOCAP dataset. The proportion of hard-negative samples indicates the ratio of hard-negative samples to all negative samples. When $\omega=0$, all negative samples are considered as easy-negative samples.}
\label{window_size_overall}
\vspace{-1em}
\end{figure}

To explore how emotional inertia affects emotion recognition in conversation, we analyze this phenomenon in the IEMOCAP dataset.
We first define hard-negative set ($\mathcal{N}_i^{\text{hard}}$) and easy-negative set ($\mathcal{N}_i^{\text{easy}}$) using different window sizes $\omega$. As illustrated in Figure~\ref{window_size_overall}, the proportion of hard-negative set among all negative samples increases rapidly with larger window sizes, indicating that more negative samples are identified as hard-negative samples.
Simultaneously, we evaluate the average cosine similarity between hard-negative and easy-negative pairs, respectively. The results demonstrate that across all window size configurations, hard-negative samples consistently exhibit significantly higher average similarity compared to easy-negative samples (all p-values $\leq 0.001$). Notably, the similarity gap is most pronounced at the smallest window size ($\omega$ = 1), where the average similarity of hard-negative samples surpasses that of easy-negative samples by a substantial margin of 0.3645, and even the positive samples.  This validates our hypothesis, making these inertia-affected utterances more challenging to distinguish in the typical MERC methods.
\vspace{-0.8em}
\subsection{Impact of Dynamic Inertia Window}
\begin{figure}[htbp]
\centering
\includegraphics[width=0.48\textwidth]{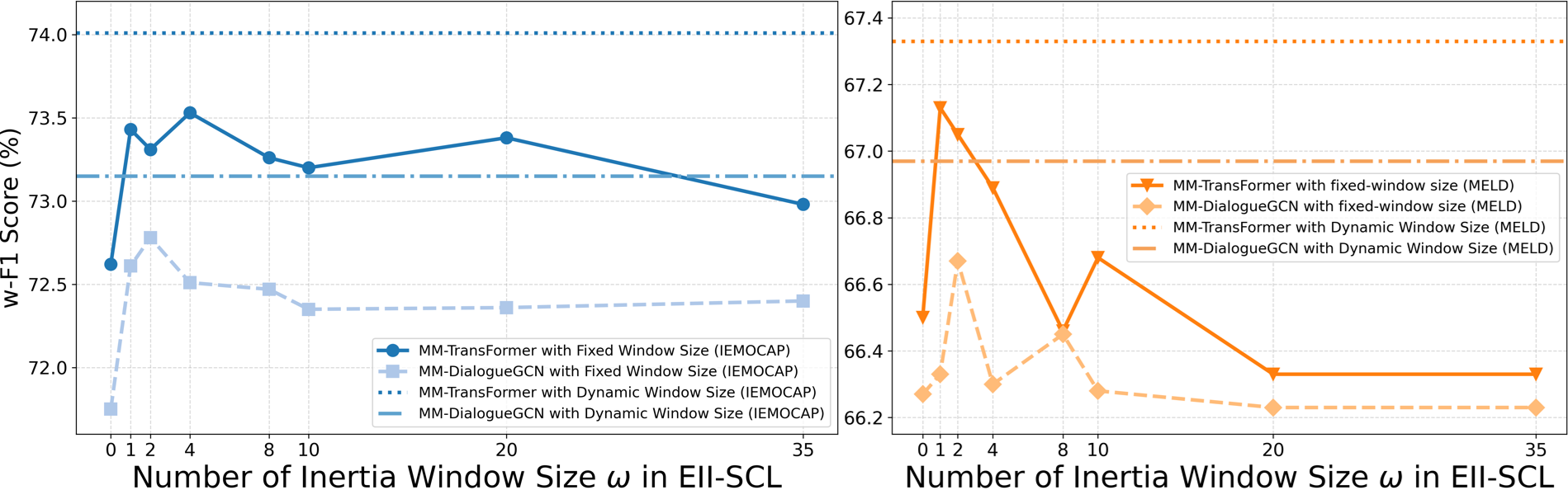} 
\caption{Impact of the dynamic and fixed inertia window size on ERC performance. The dynamic window adjusts to capture emotional persistence for each utterance.}
\label{window_size}
\vspace{-2.5em}
\end{figure}

The results in Figure \ref{window_size} compare the proposed dynamic inertia window with fixed window settings. Fixed windows introduce a predefined temporal scope that may not accurately reflect the varying persistence of emotional states in conversations. In contrast, the proposed attention-based dynamic window size $\omega^i$ adaptively adjusts the contextual range according to the semantic relevance between utterances. This adaptive mechanism enables the model to capture emotional inertia more effectively by expanding the window in emotionally stable regions while contracting it during abrupt emotional transitions. As a result, the dynamic window consistently provides more informative hard-negative samples, leading to improved ERC performance compared with fixed window configurations.

\subsection{Performance on Ambiguous Emotion Pairs}
The results in Table~\ref{tab:misclass} indicate that the proposed EII-SCL module effectively reduces misclassification rates between ambiguous emotion pairs (e.g., \textit{Hap-Exc} means the true “\textit{Happy}” label was incorrectly predicted as “\textit{Excited}”), thereby improving the overall emotion prediction performance. This demonstrates EII-SCL's capability to enhance emotion discriminability by learning more separable representations in confusing scenarios.

\begin{table}[htbp]
    \setlength{\tabcolsep} {0.77mm}
    \centering
    \vspace{-1em}
    \caption{Misclassification rates reduction of EII-SCL between ambiguous emotion pairs on IEMOCAP (\textbf{Lower the better}).}
    \begin{tabular}{c|c|c|c|c|c}
    \hline
    {\textbf{Methods}} & \textit{Exc-Hap} & \textit{Sad-Fru} & \textit{Neu-Fru} & \textit{Ang-Fru} & \textit{Fru-Neu} \\ \hline
    MM-DialogGCN & 8.85 & \textbf{19.23} & 19.56 & 27.37 & 12.22 \\
    -with EII-SCL & \textbf{6.36} & 19.88 & \textbf{14.08} & \textbf{24.54} & \textbf{11.92}  \\  \hline
    MM-TransFormer & 7.54 & 24.62 & 16.53 & 19.55 & 8.98 \\
    -with EII-SCL & \textbf{5.37} & \textbf{19.05} & \textbf{14.78} & \textbf{17.35} & \textbf{8.11} \\ \hline
    \end{tabular}
    \label{tab:misclass}
    \vspace{-\baselineskip}

\end{table}
\vspace{-1em}
\section{Conclusions} \label{conclusions}
\vspace{-0.5em}
We have proposed the emotional inertia-informed supervised contrastive learning (EII-SCL), which leverages emotional inertia—the psychological phenomenon in which emotional states tend to resist change over time—to enhance multimodal emotion recognition in conversation. Specifically, we select hard-negative samples within speaker-specific temporal windows in the supervised contrastive learning, enabling the model to learn more discriminative feature representations. EII-SCL can be seamlessly integrated into the backbone model to perceive the inherent emotional inertia in conversations, demonstrating state-of-the-art performance on both the MELD and IEMOCAP datasets. 

\section{Acknowledgment} \label{Acknowledgment}
The work presented in this article is supported by the Research Platform for Advanced Audio and Speech Signal Processing (P0049192) funded by Innovation Technology Co. Ltd.

\section{Use of Generative AI Disclosure}
\label{GenAI}
Generative AI tools were used only for language polishing and formatting assistance. All scientific content, experiments and analyses were produced and verified by the authors. 

\bibliographystyle{IEEEtran}
\bibliography{EII-SCL}

\end{document}